\documentclass[fleqn,10pt]{wlscirep}
\usepackage[utf8]{inputenc}
\usepackage[T1]{fontenc}
\usepackage{amsmath,amsfonts,amssymb}
\usepackage{graphicx}
\usepackage[colorlinks=true, allcolors=blue]{hyperref}
\usepackage[table]{xcolor}
\usepackage{relsize,setspace}  
\usepackage{url}               
\usepackage{geometry,lipsum,adjustbox,graphicx,float,gensymb,cancel,mathrsfs,url,caption,subcaption,cancel,amsfonts,textcomp,tikz,xcolor,wrapfig,,sidecap,pdfpages,fancyhdr,booktabs,multirow,capt-of,tabularx,changepage,transparent,etoolbox,titlesec,rotating,stfloats,pdflscape,ltablex,mdframed,tikzpagenodes,amsmath,color}

\newcommand{\red}{\textcolor[rgb]{0.00,0.00,0.00}}
\newcommand{\redd}{\textcolor[rgb]{0.00,0.00,0.00}}

\title{\red{Spatial dealiasing of classical geomagnetic survey data through use of a microfabricated wearable quantum magnetometer}}

\author[1,*]{Stirling Scholes}
\author[2]{Alissa Forsythe}
\author[1]{Courtney Dyer}
\author[3]{Amy Gilligan}
\author[2]{Karen Lythgoe}
\author[4]{Jenny Jenkins}
\author[1]{Marcin Mrozowski}
\author[2]{Jack-Andrew Smith}
\author[1]{Stuart Ingleby}
\affil[1]{Department of Physics, SUPA, University of Strathclyde, Glasgow G4 0NG, Scotland}
\affil[2]{School of Geosciences, University of Edinburgh, Edinburgh, EH9 3FE, Scotland}
\affil[3]{Department of Geology and Geophysics, University of Aberdeen, Aberdeen, AB24 3FX, Scotland}
\affil[4]{Department of Earth Sciences, Durham University, Durham, DH1 3LE, England}
\affil[*]{s.scholes@strath.ac.uk}

\begin{abstract}
Geomagnetic surveys provide insight into the subsurface for a range of applications, from fundamental understanding of geological processes, to mineral exploration and locating unexploded ordnance. A persistent challenge in performing such geomagnetic surveys is the joint problem of anthropogenic noise rejection and spatial aliasing, where the limited bandwidth ($<$ 10 Hz) of traditional surveying instruments introduces artefacts into the surveyed field. Optically Pumped Magnetometers (OPMs) exploit quantum mechanical effects to achieve highly sensitive and stable magnetic field measurements at comparatively high bandwidths. Recent advances in manufacturing have enabled OPMs to be packaged in compact and lightweight systems ($\sim$ 1kg), that are ideal for geomagnetic surveying. Here, we show how an OPM can directly contribute to the reduction of spatial aliasing in traditional PPM data. We carry both a PPM and OPM over a 20 km long transect across the Highland Boundary Fault (HBF) in Scotland. \redd{We} leverage the continuous acquisition of the OPM \redd{sampling at 90 Hz, equivalent to every} $\sim$1 cm at walking pace (1 m.s$^{-1}$) versus every $\sim$200 m for \redd{our PPM (which had to be stationary for measurements)} to reject magnetic noise and identify new small-scale ($<$ 200 m) geological structures. Further, we discuss the logistical advantages of the hybrid survey in terms of portability, survey delivery, data density, and data quality.

\end{abstract}
\begin{document}

\flushbottom
\maketitle
%
%
\thispagestyle{empty}

\section*{Introduction}
Magnetically susceptible buried targets acquire induced magnetization from Earth's magnetic field, generating their own secondary magnetic fields. These add to the Earth's field, appearing as magnetic anomalies relative to the large-scale spatial variability of Earth's background field. Magnetic surveying is a commonly employed technique to explore the subsurface for both geological (e.g. igneous intrusions/mineral deposits) and man-made targets (e.g. unexploded ordinance). Large-scale surveying is commonly carried out by air, however this results in limited transverse resolution due to increased distance from the source and fast travel speed. \red{Land-based surveying is required for high resolution mapping of smaller scale-targets.} Land surveys require instruments that are light, portable and weatherproof, so they can be carried by operators or attached to remotely operated vehicles, such as drones~\cite{perikleous2024application,yu2025unmanned,accomando2025advances}.\\
\\
 \red{Among the numerous instruments available for land-based geomagnetic surveys,} scalar proton precession magnetometers (PPMs) are commonly used, due to their comparatively low-cost, simple robust design, relatively high precision of $\sim$ 0.1 nT (relative to targets) and ability to capture (instrument) drift-free absolute measurements of total field strength. Despite the many advantages of PPMs, data sampling and spatial resolution is limited by the time required for a measurement. \red{This time ranges from several seconds in the case of PPMs optimised for discrete sensing, to hundreds of milliseconds in the case of PPMs employing the Overhauser effect (permitting a bandwidth of a few Hz)}. In large scale surveys of geological targets these practical limitations restrict the spacing between measurements, \red{potentially causing aliasing of small-scale anomalies while also allowing higher frequency noise sources to degrade the data.}\\ 
\\
Optically Pumped Magnetometers (OPMs) are highly sensitive magnetometers that use light to pump alkali atoms into polarised ground-states which are sensitive to external magnetic fields. When polarised atoms interact with a magnetic field they begin precessing at the Larmor frequency~\cite{Bloom_1962, OpticalReview}. This interaction provides a near instantaneous response to variations of the magnetic field, with the magnitude of the field determined using a defined constant and the measured Larmor frequency. Laser-driven OPMs have evolved from large laboratory experiments to compact and portable sensors that operate in Earth's field, i.e., without magnetic shielding~\cite{Weis_2006, Mhaskar_2012, Ingleby2022digital}. The development of \redd{Micro-ElectroMechanical System (MEMS)} cells has allowed sensitive, miniaturised OPMs to be realized, and utilizing light narrowing effects, the sensitivity can be increased~\cite{Shah_2013, Appelt_1999, Scholtes_2011}. These developments have led to new and exciting applications for OPMs including biomedical imaging ~\cite{Hill_2020, Feys_2023}, space weather ~\cite{Mrozowski_2024}, GPS denied navigation ~\cite{Canciani_2022}, underwater surveying ~\cite{Page_2021}, and geomagnetic surveying ~\cite{yu2025unmanned}.\\ 
\begin{figure}[b!]
\begin{center}
  \includegraphics[width=0.9\linewidth]{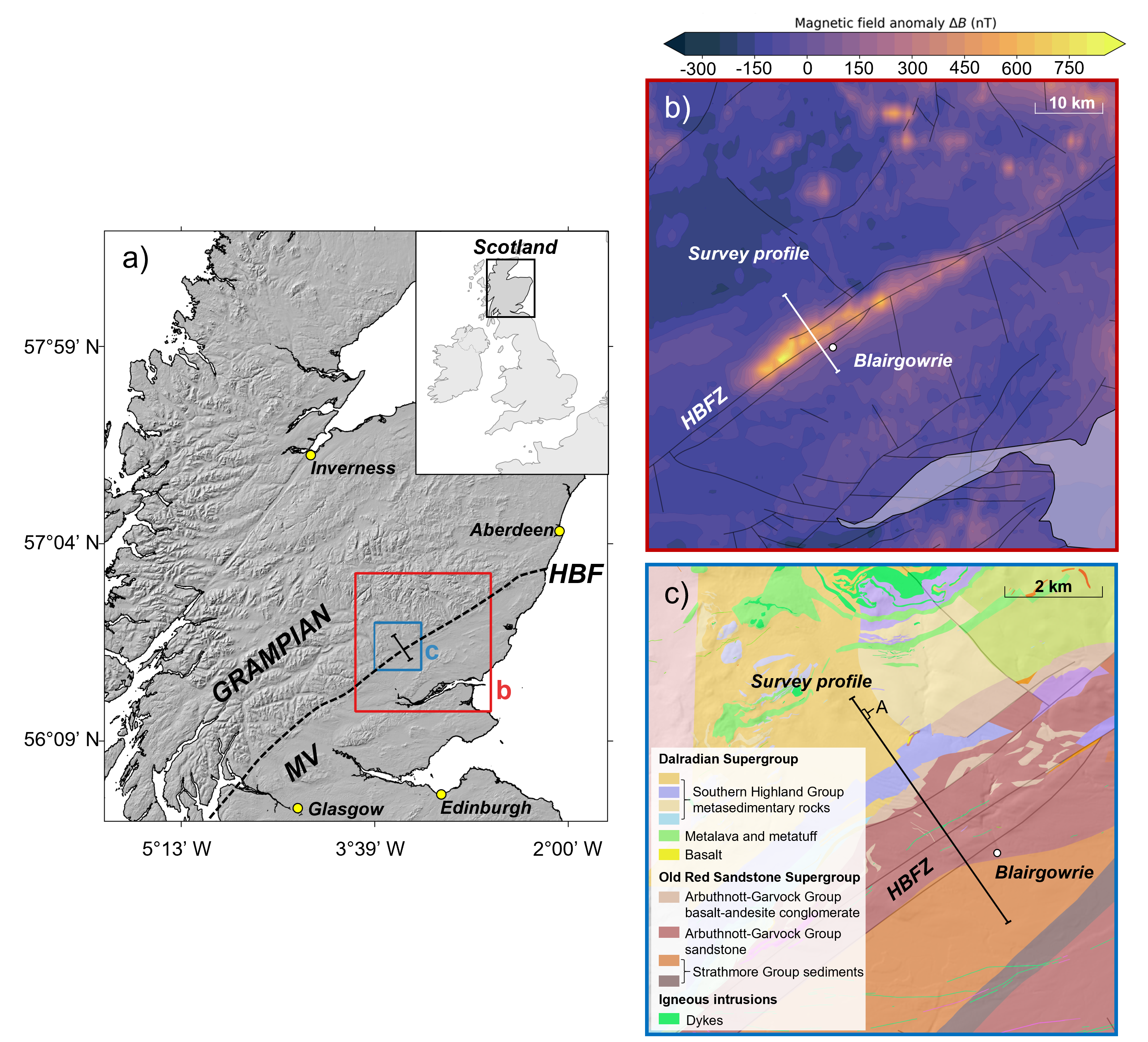}
  \caption{a) Map of the study area across the Highland Boundary Fault (HBF) in Scotland, separating the Grampian terrane to the north from the Midland Valley (MV) terrane to the south. Inset shows this location within the United Kingdom and Northern Ireland. The solid black line marks the NW-SE trending survey profile through Blairgowrie. The red and blue squares correspond to the areas shown in b) and c), respectively. b) Magnetic field anomaly map from~\cite{bgs_mag} of the study area showing the >600~nT Blairgowrie anomaly. c) Geological map of the Highland Boundary Fault zone (HBFZ) and surrounding area. A: location of the small-scale modelling in Figure~\ref{fig:spike_model}. Contains British Geological Survey materials © UKRI 2025.}
  \label{fig:study_area_map}
\end{center}
\end{figure}\\
In this study, a recently developed compact OPM, employing the double resonance technique~\cite{Brossel_1952, Bloom_1962, Weis_2006}, is used. This OPM is portable and continuously acquires data, \red{possessing a mean sampling rate of 90 Hz, and thus a significantly higher spatial resolution and noise tolerance than available PPMs. This allows the OPM to resolve more localised anomalies. Further, the compact and lightweight ($\sim$ 1 kg total weight) nature of the sensor enables a wearable configuration. A modified utility vest is used to house the OPM system enabling hands-free operation.} The OPM uses the well-defined gyromagnetic ratio associated with the $^{133}$Cs ground state (F=4) and is capable of a precision of 3 pT at 1 Hz ~\cite{Mrozowski_2024}.\\ 
\\
\red{Here, we pair the researcher-developed microfabricated double-resonance OPM with a Consumer Off-The-Shelf (COTS) PPM.} Our field site is across the Highland Boundary Fault (HBF), one of Britain’s major geological terrane boundaries, where there is a pronounced magnetic anomaly of interest (Figure~\ref{fig:study_area_map}). The HBF zone extends more than 250~km across Scotland and separates two distinct geological terranes: (i) the Neoproterozoic metasedimentary rocks of the Grampian Highlands to the North West and (ii) the Devonian to Carboniferous sedimentary and volcanic successions of the Midland Valley to the South East. Despite detailed surface investigations, the deeper structure of the HBF remains the subject of long-standing debate. Some authors suggest that it is a major crustal suture \cite{CURRY1982, BLUCK1984, BLUCK2010}, while others argue that it is merely a shallow zone of younger, overprinting deformation \cite{HENDERSON_ROBERTSON1982, TANNER_SUTHERLAND2007, TANNER2008}. A prominent (>600~nT) magnetic anomaly, trending NE-SW along the HBF for over 40~km can be seen in the Great Britain Aeromagnetic Survey \cite{bgs_mag} in the vicinity of the town of Blairgowrie, Perth and Kinross, Scotland (Figure~\ref{fig:study_area_map}b). Its position coincides with a structurally complex segment of the HBF zone (Figure~\ref{fig:study_area_map}c), comprising a succession of faulted blocks, basaltic-andesitic lava flows, igneous dykes, and several serpentinite outcrops \cite{CRANE2002}. A detailed magnetic ground survey following roads in the area was previously conducted \cite{FARQUHARSON1992} and the results were interpreted as a long-wavelength magnetised body at depths of 2~km to at least 13~km, consistent with serpentinite inferred from nearby field observations. Shallower anomalies were attributed to intrusions of andesitic lavas, consistent with geological mapping. For our study, an approximately 20~km NW-SE trending profile through the Blairgowrie region was chosen, located to the west of the historic transect.\\
\\
\red{We present the modelling of new spatially small geomagnetic anomalies, the detection of which is directly facilitated by simultaneously mapping the HBF using both a PPM and an OPM. Specifically, we leverage the sensitive continuous acquisition of the OPM to monitor the magnetic field in the area around the PPM sample points, identifying PPM sample points degraded by anthropogenic magnetic clutter, while retaining the absolute accuracy of the PPM in non-degraded samples. Together, the two sensors realise a system capable of achieving an accuracy and spatial resolution greater than either sensor could achieve individually without meaningfully increasing survey time. Further, we discuss the data quality and survey efficiency enabled by our approach. } 


\section*{Results}
\begin{figure}[t!]
    \centering
    \begin{tabular}{c c}
        \includegraphics[width=0.41\linewidth]{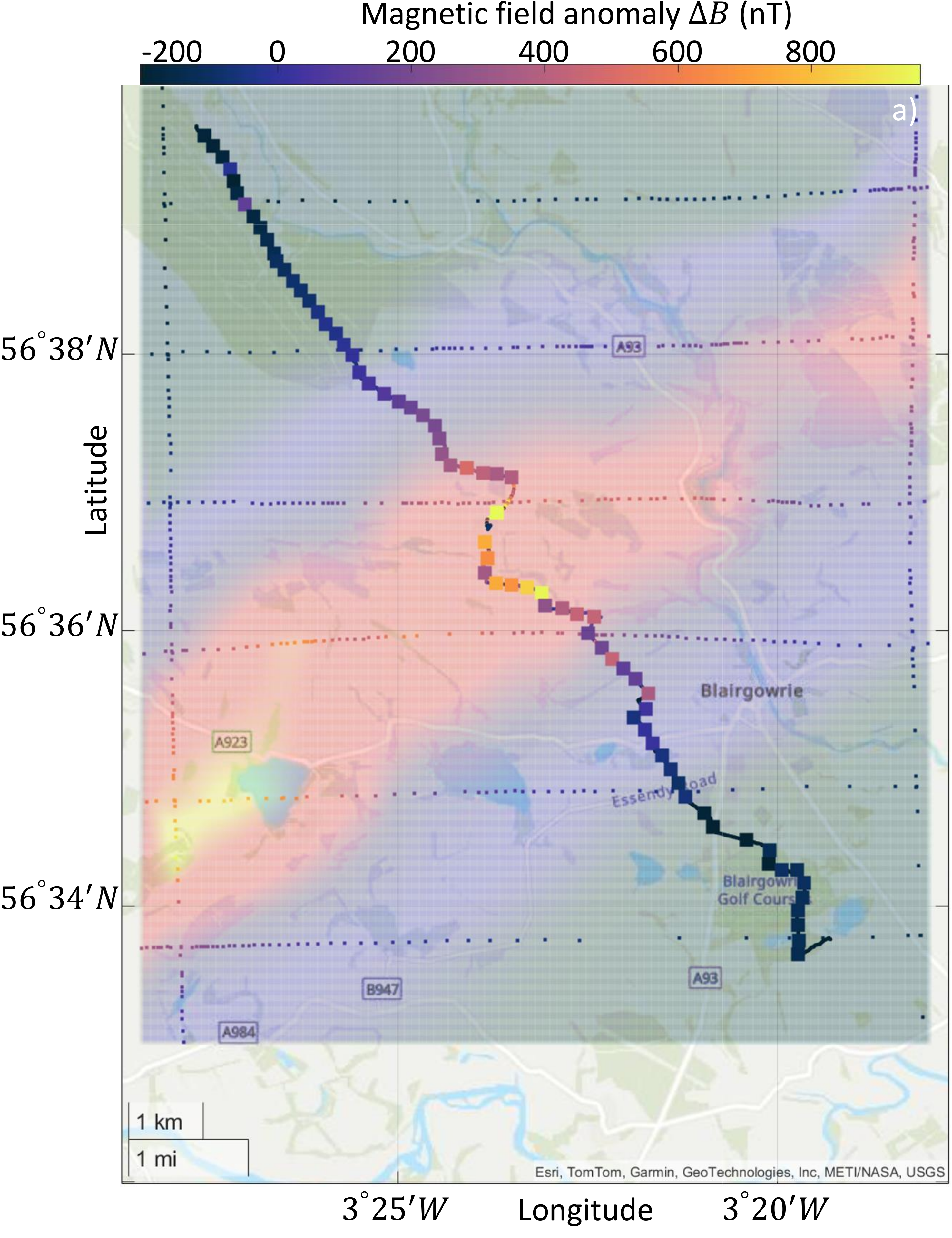}&
        \includegraphics[width=0.41\linewidth]{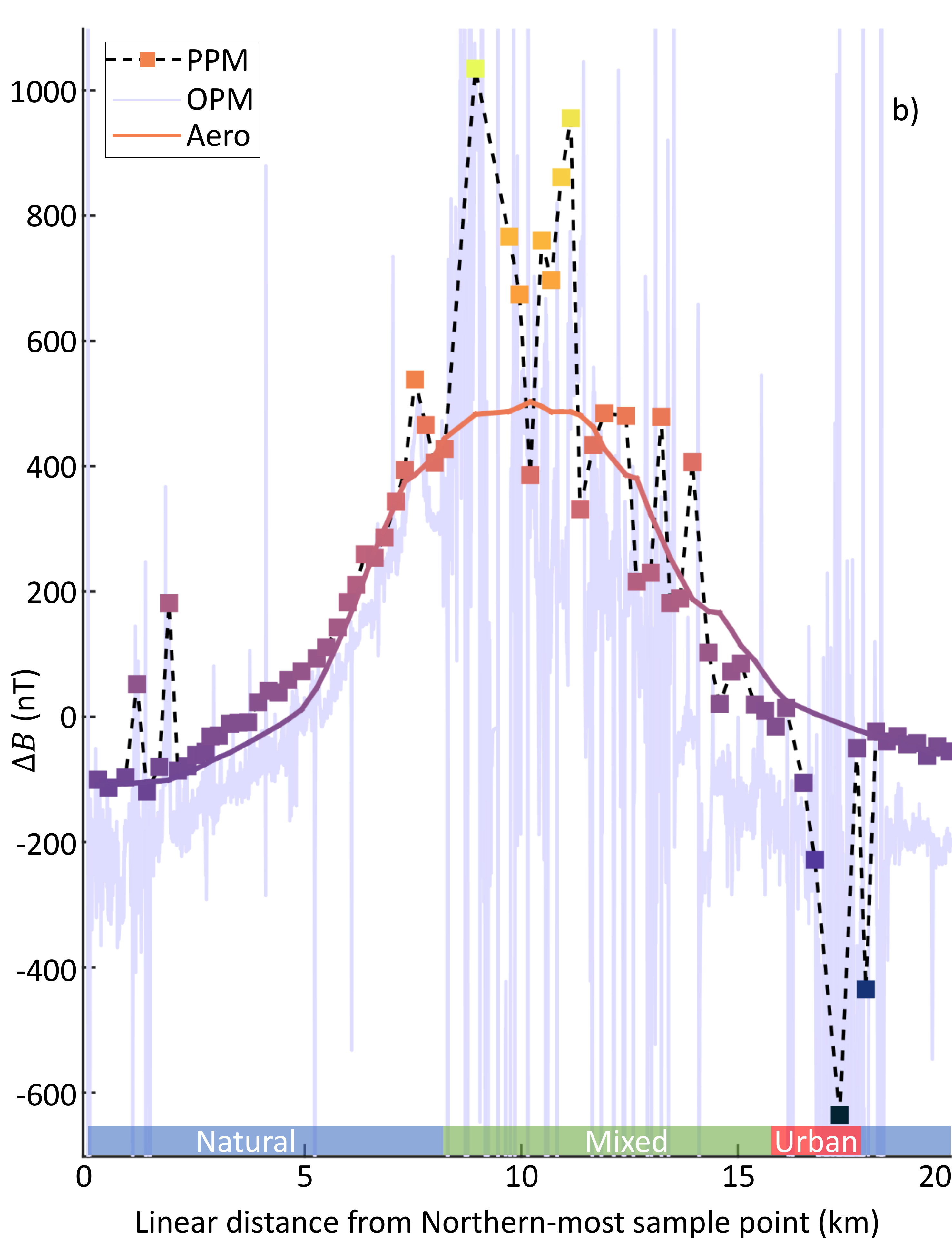}\\
    \end{tabular}
    \captionof{figure}{a) Map of the PPM (discrete squares) and OPM (continuous line) anomaly values $\Delta B$ along the survey path. The survey path transects the HBF as shown in Figure~\ref{fig:study_area_map}. The overlay is interpolated from the BGS aeromagnetic anomaly survey\cite{bgs_mag} with the aeromagnetic survey points plotted as small dots. b) Data plotted as a linear path from the Northern-most sample point. The values of the Aeromagnetic survey \cite{bgs_mag }are represented by the solid line of variable colour. The labels on the x-axis indicate land usage for different segments of the survey: ``natural" = land with minimal magnetic clutter, ``mixed" = mixed use, e.g. farm land, and ``Urban" = populated regions. Contains British Geological Survey materials © UKRI 2025.}
    \label{fig:PPM_OPM_Aero}
\end{figure}
Magnetic data was collected with both PPM and OPM instruments over two days in June 2025 (see Supplementary material section 1 for exemplar sensor operation). The survey follows footpaths and roads, along a $\sim$20 km long NW-SE transect, perpendicular to the NE-SW orientation of the Highland Boundary Fault (Figure~\ref{fig:PPM_OPM_Aero} a). Instrument operators began the survey each day in the centre of the transect. Both OPM and PPM operators started at the same time and walked along the same path. The OPM team was able to cover a larger distance (e.g 16 km vs 12 km on Day 1) in the same time due to the OPM's near continuous data collection. The upper limit on the OPM system's range is set by battery capacity ($\sim$ 9 hours). \red{Here, we show results only in areas common to both instruments for direct comparison.}\\ 
\\
Figure~\ref{fig:PPM_OPM_Aero} a) shows a map of the acquired PPM and OPM data compared to the historical aeromagnetic survey data~\cite{bgs_mag}. PPM values are shown as discrete squares, while the OPM data is shown as a continuous line. For details on how the anomaly values were calculated see the Methods Section. The transparent overlay is interpolated from the BGS aeromagnetic anomaly survey, with the aeromagnetic survey points shown as small dots.  Qualitatively, good agreement can be seen between our survey data and the long-wavelength lower-resolution aeromagnetic data. Figure~\ref{fig:PPM_OPM_Aero} b) plots the data against a linear path from the northernmost sample point. The interpolated aeromagnetic values in Fig~\ref{fig:PPM_OPM_Aero} b) resemble a smoothed version of the measured PPM data. This is consistent with the greater sampling density of the PPM survey ($\sim$ every 200 m) and the altitude difference between the surveys. The OPM data in Figure~\ref{fig:PPM_OPM_Aero} b) augments the PPM data, with its significantly higher transverse resolution, although absolute values of magnetic anomaly are offset from the PPM and aeromagnetic data.


\subsection*{Outlier identification of magnetic data by combining PPM and OPM measurements}
\begin{figure}[h!t!]
    \centering
    \begin{tabular}{c}
        \includegraphics[width=0.9\linewidth]{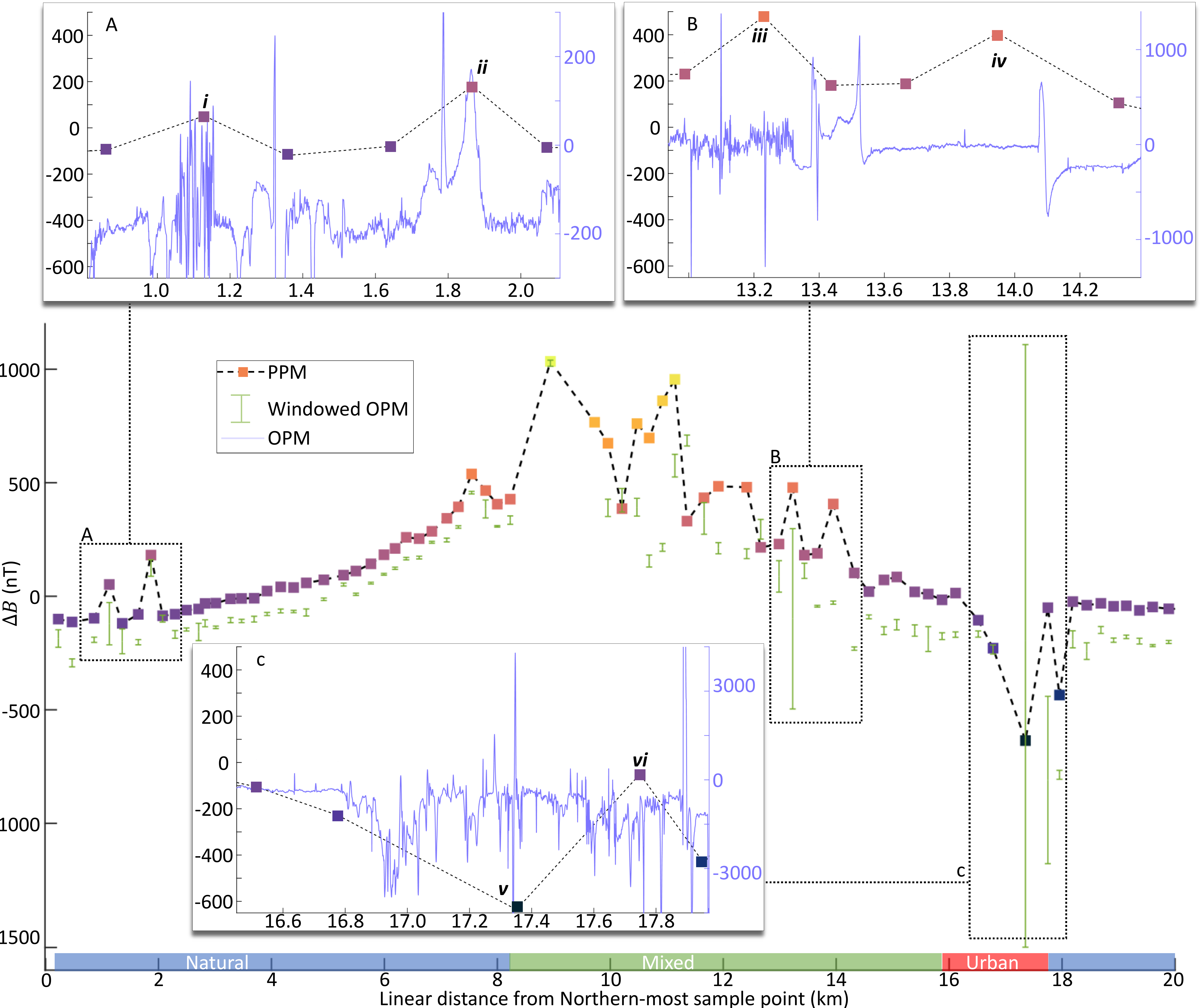}\\
    \end{tabular}
    \captionof{figure}{N-S profile of PPM and OPM data. PPM data is shown as coloured squares, connected by a black dashed line. OPM data is shown in green, with  error bars showing the standard deviation in a 20 m window around each PPM sample point. The insets marked A, B, and C correspond to the boxed regions in the main plot. Within each inset, the OPM data is the blue line, while the PPM data is the same as the main figure. \textit{i - vi} represent PPM points of interest within each inset.}
    \label{fig:OPM_Std_plot}
\end{figure}
In Figure~\ref{fig:PPM_OPM_Aero}, from 0-8 km, the OPM data changes with distance in a relatively smooth and consistent manner, reflecting the low noise rural environment. This is corroborated by the similarity in trend between the OPM and PPM data. Several peaks observed in the PPM data in this section also occur in the OPM data, suggesting robust changes in the field strength on the order of 100s of meters (for example see inset A in Figure~\ref{fig:OPM_Std_plot}). By contrast, the sharp peaks in the OPM data without corresponding changes in PPM values, likely represent local small-scale anthropogenic sources, for example metallic gates along the footpath. Between 8 and 18 km, the survey passed through mixed use and urban areas, significantly increasing the noise measured by both the OPM and PPM instruments.\\
\\
To identify PPM sample points affected by magnetic noise, the OPM data was divided into 20 m windows centred around each PPM sample point. The standard deviation of measured OPM values within each window provides an indication of the impact of magnetic noise on the PPM sample, i.e., a large standard deviation indicates a large change in magnetic field in close proximity to the PPM sample point, characteristic of anthropogenic noise. \red{The size of the window is related to the spacial scale of the observed anthropogenic noise sources. Specifically, sharp anomalies (for example from fences) and larger dipolar anomalies span tens of meters (inset A and B in Figure~\ref{fig:OPM_Std_plot}) with geological anomalies spanning hundreds of meters. Twenty meter windows were selected to reliably identify noise sources without falsely encompassing the small-scale geological features of interest.} The main panel of Figure~\ref{fig:OPM_Std_plot} shows the PPM sample points as coloured square markers on a black-dashed line. For each PPM point the windowed OPM data is shown by green bars. Specifically, each green bar is centred around the mean value of the OPM data in the 20 m window with the extent of the bars indicating the standard deviation of OPM values in the window. The insets of Figure~\ref{fig:OPM_Std_plot} correspond to the boxed regions in the main plot. Within each inset, the OPM data is the blue line (right blue axis), while the PPM data is the same as the main figure (left axis).\\
\\
Examining the PPM data in inset A of Figure~\ref{fig:OPM_Std_plot},  two positive peaks, \textit{i} and \textit{ii}, occur at ~1.1 km and 1.9 km respectively. The OPM data also has a positive peak at around 1.9 km, with a small standard deviation, indicating that this peak is an anomaly representing a sub-surface geological source. By contrast, the OPM data in the vicinity of the positive PPM peak \textit{i} at 1.1 km shows rapid oscillation, resulting in the larger standard deviation. This indicates the PPM point \textit{i} near 1.1 km is likely from an anthropogenic source, and should not be considered in geological interpretations.
\\
\\
The points examined in window B further corroborate the benefits of the tandem PPM-OPM approach. Window B contains two peaks \textit{iii} and \textit{iv} (at 13.2 and 14 km) in the PPM data, of a similar magnitude to the those seen in window A. The OPM data for point \textit{iii} shows that the PPM measurement was taken close to a noise source as indicated by the sharp dip in the OPM data near 13.2 km, and the correspondingly large standard deviation. This suggests this PPM data point should not be considered in geological modelling.  The positive PPM peak \textit{iv} near 14 km has a small standard deviation, however does not have a corresponding peak in the OPM data, unlike peak \textit{ii} at 1.9 km in window A. Therefore, this can likely be attributed to a transient noise source. Although the OPM and PPM surveys were conducted on the same day, the OPM survey team passed through the mixed use agricultural area $>$ 1 hour before the PPM team due to faster data collection, meaning mobile noise sources, such as parked vehicles, may differ between the two passes. Further, the asymmetric dipolar feature near 14.1 km in the OPM data is characteristic of moving magnetic sources.\\
\\
Window C contains the largest observed OPM standard deviation. The OPM data near 17.4 km show a large change in value over a short distance ($\approx$ 12 m) located within 5 m of the PPM sample location. This PPM sample point, \textit{v}, was taken in close proximity to an electrical utility pole (as verified by Google Street View). Similarly, the PPM sample, \textit{vi}, near 17.8 km was taken in close proximity to an observed OPM spike, justifying removal of these points from geological modelling. This approach, of combining both the OPM and PPM data, allows for a justifiable and robust method to identify points that are outliers, which can then be removed prior to modelling long wavelength magnetic structure (Supplementary Material Section 2). Further, it allows for points that may otherwise have been discarded as outliers (such as the peak \textit{ii} at 1.9 km in window A), to be identified as points of interest for further modelling and interpretation. 

\subsection*{Detailed modelling of small-scale geological structures}\label{modelling_section}
\begin{figure}[t!]
\begin{center}
  \includegraphics[width=0.8\linewidth]{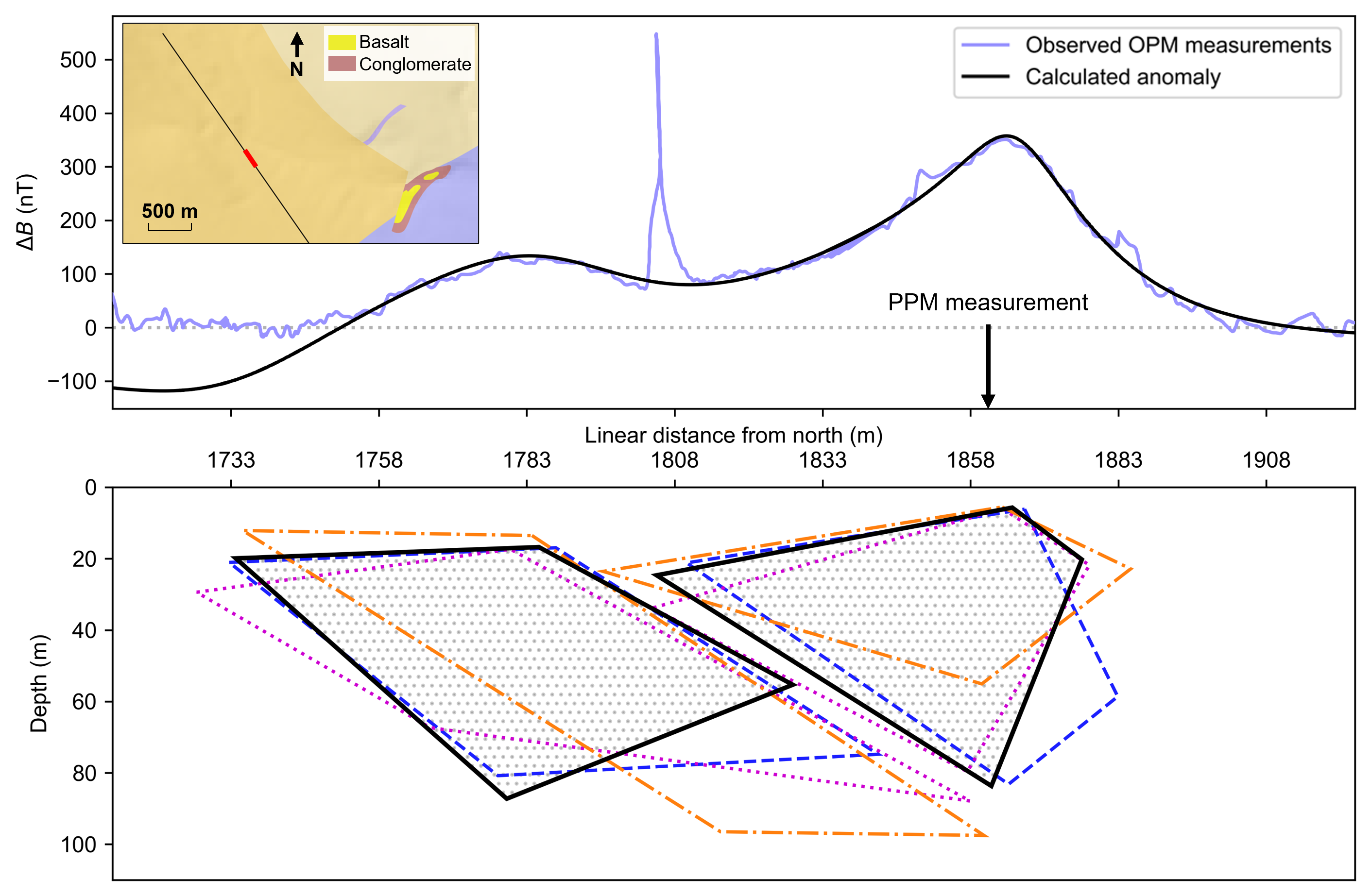}
  \caption{Forward modelling results for anomalies in the OPM data between 1.75 and 1.90~km along the survey profile (see Figure~\ref{fig:OPM_Std_plot}). The top panel shows the observed OPM measurements (purple) and \red{the fit derived from the calculated anomaly (black). The arrow indicates the position of the singular PPM measurement related to this anomaly. The inset highlights the modelled section (red) on top of the geological map. The bottom panel shows the cross-sectional geometries for a set of indistinguishable solutions generated by the forward modelling. The solution corresponding to the fit shown in the upper panel is shown by the black-outlined shaded bodies. The pink, blue, and orange cross-sections show other solutions which produce equivalent fits. The magnetic susceptibility of all modelled bodies is 0.025 (SI units).}}
  \label{fig:spike_model}
\end{center}
\end{figure}
The higher spatial sampling in the OPM data permits smaller scale structures to be resolved in comparison to PPM data. One example (around point \textit{ii} of inset A of Figure~\ref{fig:OPM_Std_plot}) is shown in detail in Figure~\ref{fig:spike_model}, from  1.7 and 1.9~km along the survey profile. In the PPM data, an $\approx$ 180~nT positive anomaly is seen at 1.9 km. The PPM data consists of only a single sample point that could easily have been considered an outlier, given the smooth decreasing trend observed surrounding it. The OPM data gives confidence that this is likely an anomaly due to a sub-surface geological source, revealing greater complexity (two short-wavelength anomalies), and due to the significantly higher sampling of the OPM data, means we can model this section of the transect in detail, which would not be possible from just the PPM data. The very short wavelength spike in the OPM data at 1.8 km is likely anthropogenic noise, and is disregarded in the modelling.\\
\\
The short-wavelength anomalies suggest the presence of two small, shallow magnetic sources. Considering the geological map, these were modelled as igneous bodies, possibly extending northwards from the basalt outcrop just to the SE of this section of the profile. Similar basaltic andesite flows have been traced along the river Ericht (east of the profile; Figure~\ref{fig:PPM_OPM_Aero}), interbedded with decomposed hematitic lava, which are associated with relatively high magnetic susceptibilities \cite{CRANE2002}. Furthermore, small outcrops of lavas at a similar stratigraphic level have been observed in the wider area (not shown on the geological map), with reports noting up to five distinct flows, each approximately 10-15~m thick \cite{CRANE2002}. Collectively, these observations support modelling the sources as shallow igneous bodies extending to depths up to 100~m below the surface. Magnetic susceptibilities were varied within the range reported for local lava units \cite{FARQUHARSON1992} and were ultimately fixed at 0.025 for both bodies. The sources were initially modelled as vertically oriented rectangular prisms; however, introducing a dip was required to reproduce the shape of the observed anomaly without considering induced magnetisation. This may be consistent with dipping units within the Tay Nappe \cite{krabbendam1997}, generally towards the SE or NW depending on fold limb, however an exact dip and dip direction of this section on our study profile has not been recorded. Both bodies were assigned an along-strike extent of 500~m, although this parameter is poorly constrained and greater lengths seemed to have minimal influence on the model response.\\ 
\\
A final model is shown in Figure~\ref{fig:spike_model}, comprising two shallow, southward-dipping magnetic bodies (black shaded polygons).
\red{Due to the inherent non-uniqueness of magnetic forward modelling, we provide a set of possible solutions which yield an equivalent measured anomaly. The cross-sections of these bodies are shown as blue, pink, and orange polygons in Figure~\ref{fig:spike_model}.}
Note that changes to one body require compensating adjustments to the neighbouring body due to the superposition of their magnetic signatures. However, \red{the large degree of intersection between the possible solutions to the model clearly demonstrates the advantage of using continuous OPM data over discrete PPM measurements when investigating small-scale geological structures.} The short-wavelength anomalies could not be resolved using the PPM due to spatial aliasing (see Figure~\ref{fig:OPM_Std_plot}), and it would not have been possible to model the subsurface on this short length scale from a single point in the PPM data. Even if the point had been included in larger scale modelling, at best modelling would likely have resulted in  modelling a single broader source.



\section*{Discussion}
Logistically, the OPM instrument has shown significant practical advantages in data collection. The OPM generated orders of magnitude more measurements and covered equivalent distances in less time (typically saving multiple hours per day) due to its high sampling rate relative to walking pace. In comparison, the several seconds taken for a PPM measurement, along with the necessity for the operator to keep the instrument static while a measurement is made, significantly impacts the speed of data collection. The light weight of the OPM (which has been designed to be carried in a wearable vest), means the distance covered is effectively only limited by practicalities such as walking speed and battery capacity. Given this, and the high sampling rate relative to walking pace, future developments are intended to explore practicalities of mounting instrumentation on semi-autonomous platforms such as drones, which would further enhance the spatial coverage possible.\\  
\\
\red{The M$_x$ OPM used is a low-drift\cite{Mrozowski_2024} relative instrument and is used to best advantage in conjunction with an absolute magnetometer, such as the PPM used in this study. Current developments in OPM design offer increasingly accurate and stable configurations which will continue to enhance field-able instrumentation\cite{Hunter:22, Hunter:23, HUNTER_2025}.} The absolute offset between the OPM data and the PPM data can be attributed to a combination of sensor and environmental dynamics, making it non-constant. In magnetically quiet environments (marked 'natural' in Figs.~\ref{fig:PPM_OPM_Aero} b) and~\ref{fig:OPM_Std_plot}), the offset can be attributed to known small systematic offsets found in laser-pumped alkali magnetometers, such as light-shifts and heading errors \cite{chalupczak2010}, as evidenced by the approximate stable scale in the static shift between datasets of 91$\pm$33 nT. In mixed and urban environments, the static offset is combined with dynamic noise sources. This is supported by the high variability in the scale of the shift observed in the mixed and urban regions, where the survey transect crosses through the town of Blairgowrie (264$\pm$378 nT). Beyond 18 km the transect returns to a magnetically quiet area, where values of the shift become more consistent again (152$\pm$26 nT), and are comparable in standard deviation to those at $\leq$ 8 km, indicating a return to purely systematic cause.\\
\\
The data reveals magnetic anomalies on a range of scales. The long-wavelength peak observed in the aeromagnetic data is mirrored in the terrestrial data. The PPM data reveals additional details, with  shorter wave-length variations.  However, only when analysed in conjunction with very high-resolution OPM data, does it become possible to distinguish which PPM data points represent anthropogenic sources and which reveal further detail of the subsurface geology. While several data spikes are simply outliers due to man-made magnetic noise, some represent real geological features (for example \textit{ii} in inset A in Figure 3), which only become possible to model using very high resolution OPM data. This has allowed  the identification of geological features that have not been mapped previously. 

\section*{Conclusions}

Using a combination of a COTS PPM and a researcher-developed microfabricated double-resonance OPM, we have completed a magnetic anomaly survey crossing the Highland Boundary Fault. This combination of instruments yielded a richer dataset, facilitating more detailed interpretation than either in isolation. Combining the high portability, sensitivity and data rate of the OPM with the high accuracy and intrinsic calibration of the PPM has allowed us to spatially de-alias the effects of anthropogenic magnetic clutter from sparse PPM data points. This was achieved using the windowed standard deviation of the spatially dense OPM data to triage PPM data points, identifying new features of potentially geophysical origin, which would not be found from the PPM alone. PPM data in less cluttered regions allows for accurate baseline correction of the OPM data. The combination of both instruments offers a powerful new tool for geomagnetic exploration and spatial awareness. Due to the OPM's unique use of microfabrication and chip-scale components, this can be realised in a practical, mass-producible and deployable form factor.


\section*{Methods}
\label{sec:Methods}

\subsection*{PPM methodology}
A standard PPM ~\cite{packard1954ppm} (Geometrics G-857 magnetometer) capable of measuring absolute magnetic field strength with a theoretical precision of $\sim$0.1 nT was used in the study. The  sensing element of the instrument is formed of a cylindrical "bottle" of hydrocarbon fluid. A polarising current is passed through a copper solenoid surrounding the bottle to generate a  magnetic field (on the order of 10~mT), with which the moments of hydrogen protons become aligned. When the current is switched off, protons become re-aligned with the Earth's weaker background field. The process of realignment is associated with the emission of electromagnetic waves, the frequency of which are proportional to background field strength $B$, according to the equation:
$B=\frac{2\pi}{\gamma_{P}} f$
where the known constant of the gyromagnetic ratio $\gamma_{P}$ (ratio of the magnetic moment of a proton and the angular moment of its spin) is equal to an accurately known value of $2.6752219 \times 10^{8}  s^{-1}T^{-1}$  \cite{FundamentalsGeophysics}. Multiple frequency cycles are counted to ensure accurate measurements (based on an induced electrical signal in a detector solenoid around the bottle), meaning that readings take several seconds.\\  
\\
The sensing "bottle" was carried on a 2 m high pole, to isolate it from interference, and orientated vertically (following manufacturers recommendations for UK latitudes), such that the angle between the polarising field and background field was maximised. Measurements were taken a short time after the operator stopped walking to allow liquid in the sensor to settle. Operators, carried the instrument control unit in a wearable harness and wore nothing metallic, carrying no electronic devices to minimise interference. A minimum of three repeat readings were taken every 200 m along the survey line to assess data error (more if there was significant variability observed between repeats), with data location recorded on a handheld GPS, with predicted accuracy of 3-5 m. \red{The maximum bandwidth of the PPM is 0.6 Hz and the sensor has a sensitivity of 100 pT.} A continuously recording PPM base-station (sampling every 15 seconds) was left at the centre of the survey, to capture temporal variations in the background field. Repeat measurements were averaged and the standard deviation calculated prior to reducing the data from magnitude to an anomaly.

\begin{figure}[b!]
    \centering
       \includegraphics[width=0.7\linewidth]{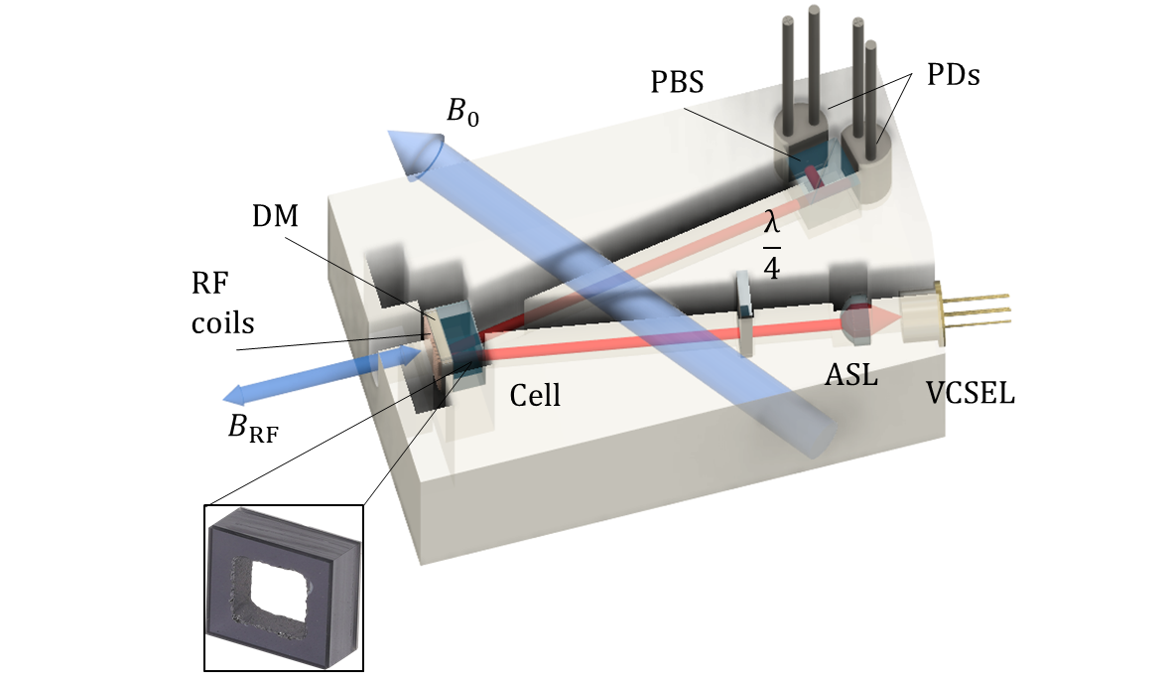}
    \captionof{figure}{Schematic of OPM sensor head. VCSEL: Vertical Cavity Surface Emitting Laser, ASL: Aspheric lens, $\lambda / 4$: Quarter-wave plate, DM: Dielectric Mirror, PBS: Polarising Beam Splitter, PDs: Photodiodes. $ B_0$ denotes the external magnetic field (in this survey, Earth’s magnetic field) and $ B_{RF} $ denotes the oscillating RF field generated by the RF coils. The angle between the incident and reflected beam is $20 ^\circ $. \red{\textit{Inset:} photograph of the optimised microfabricated cell used in this work.}
}
    \label{fig:OPM_Schematic}
\end{figure}

\subsection*{Optically pumped magnetometer methodology}

Using a compact, wearable OPM the variations in Earth’s magnetic field due to geomagnetic anomalies are measured. A circularly polarised beam optically pumps atoms into a dark state and creates a magnetization, precessing at the Larmor frequency around an external magnetic field ($B_0$). The Larmor frequency is dependent on $B_0 $ and the gyromagnetic ratio of the ground state hyperfine energy level ($\gamma$), $\omega_L = \gamma  B_0 $. The atomic population is driven into an absorbing state, using a radio frequency oscillating field ($B_1(t)$), where the precession is detected using a linearly polarised probe beam. The magnetic resonance is obtained when the resonant condition $\omega_L = \omega_{rf}$ is reached. In the M$_x$-configuration, the magnetic field $B_0$ is oriented at $45^\circ$ from the laser beam. The sensitivity of the OPM is improved by exploiting ‘light narrowing’ of magnetic resonances. The resonance linewidth broadens at increasing light intensities but narrows again due to Spin-Exchange (SE); this broadening is sufficiently exceeded by narrowing when the SE is sufficiently rapid. In this survey, Earth's magnetic field is the external magnetic field.\\ 
\\
\red{The M$_x$ OPM used in this study achieves a portable low-SWaP form factor through the novel combination of a chip-scale laser with a microfabricated alkali cell optimised for geomagnetic measurements. A single laser scheme is used, in which a Vertical Cavity Surface Emitting Laser (VCSEL), tuned close to the $^{133}$Cs D1 absorption line, both optically pumps and probes the magnetic resonance.} The laser is placed into the sensor head package, then collimated using an aspheric lens before a half wave-plate converts the linearly polarised beam to circularly polarised light. The beam enters a \red{custom-microfabricated} $^{133}$Cs vapour cell (6 mm x 6 mm x 3 mm), containing approximately 200 Torr of $N_2$ buffer gas, at an angle of 10$^\circ$ from the normal of the cell. A dielectric mirror is attached to the rear of the vapour cell, creating a double-pass configuration in which the laser leaves the cell with a total reflected angle of 20$^\circ$. Radio Frequency (RF) coils on the back surface of the dielectric mirror create a weak, oscillating RF field. These coils also function as a heating element, improving the atomic density of the cell. The light then travels through a Polarising Beam Splitter and is detected on two photodiodes in a balanced polarimeter configuration. The sensor head is depicted in Figure~\ref{fig:OPM_Schematic}.\\
\\
The OPM is used in a wearable configuration, requiring a low size, weight and power (SWaP) system. The system relies on a fanless mini-PC (Fitlet 3) for local data logging and computation. The low power requirements of the PC and sensor allow the entire system to run for several hours using a COTS 10 000 mAh portable battery bank. Longer operational times can be achieved with the use of larger power banks. To protect the sensor from the environment, the OPM sensor head is placed in an IP65 waterproof enclosure and the associated cabling fed through a conduit. The sensor head and electronics are attached to a modified utility vest using a MOLLE-PALS system with the sensor enclosure mounted in the small of the back. The electronics and sensor head are placed on opposite sides of the body to limit electronic noise in the survey data.\\
\\
During the survey, location data was logged using a recreational GPS tracker. Location was logged at a cadence of 1 Hz with interpolation later used to infer coordinates between logged points. The mean sample cadence of the OPM is 90 Hz which at an average movement speed of 1.1 $m.s^{-1}$ (4 kph) results in a spatial resolution of < 2 cm. OPM data collected in this way was first resampled to a 10 Hz standard, then a lowpass filter with a cutoff of 0.5 Hz was applied to the data. The data was lowpass filtered to remove high frequency artifacts, for instance the electrical noise from the power grid. \red{This is then equivalent to a sampling wavenumber of 9.1 m$^{-1}$. The Nyquist limit of the data is 45 Hz and 5 Hz for the raw and processed data respectively. At the average movement speed of 1.1 $m.s^{-1}$, the 5 Hz limit corresponds to a spatial bandwidth limit of 22 cm.} The same diurnal correction as was used in the PPM processing was applied to the OPM data, as detailed below. The OPM data was then subtracted from an IRGF-2025 model to calculate the magnetic anomaly.


\subsection*{Data reduction: Diurnal correction and reduction to anomaly }

The Earth's magnetic field will vary temporally due to space weather. The effect of this needs to be removed prior to calculating the magnetic anomaly. In this study, we set up a base station with an identical instrument (Geometrics G-857) to that used in the PPM survey at Middleton of Glasclune Farm (56.6125 N, -3.3966 W). Measurements of the absolute magnetic field were automatically recorded every 15 s throughout the survey period.\\
\\
On day two there were four occasions where there were large jumps in the data (3.8 nT, 6.3 nT, 6.8 nT and 27.8nT within a 15s window), which can be attributed to the movement of vehicles into or out of a parking area close to the base station. To deal with these, data from times before these spikes are shifted as appropriate. To deal with transient noise spikes, such as farm vehicles moving on a nearby road, we first remove all data points with a gradient exceeding +/-0.1 nT/s, and then apply a rolling mean to the time series. We test 1.5, 5, and 10 minute rolling means, and use the 5 minute rolling mean for the results presented here. On day one, the range was 31 nT during the course of the survey and on day two, the range was 39 nT. The gradient of the changes to the field was then calculated in nT/s for use in the diurnal correction. \red{The uncertainty associated with the base station instrument is 100 pT, significantly below the measured diurnal swing on both days. As such, experimental errors introduced by this uncertainty are taken as negligible due to their small scale relative to the size of the measured anomalies. Further, the data from our base station was compared with the measurements made at the British Geological Survey Magnetic Observatory in Eskdalemuir, where similar diurnal changes were also observed.}\\  
\\
Following Ref~\cite{milsom2013field}, we pick a standard value of 50935 nT (an approximate mid-value of those recorded at the base station). For each point on the PPM and OPM surveys,  we calculate the difference between the base station measurement at the time a measurement of the magnetic field was taken and the standard value, using the gradient to interpolate if necessary. This difference is then added to the magnetic field measurement, ranging from -37 nT to 26 nT depending on the time of the measurement, using the 5 minute rolling mean base station data. This is small compared to the anomalies of interest.\\ 
\\
Using the diurnally corrected measurements, we then calculate the anomaly relative to the IGRF model. This is done using the pyIGRF \cite{alken2021international}, for the latitude, longitude, elevation, and time of each reading.


\subsection*{Magnetic modelling}
2.5D magnetic forward modelling was carried out using the software Mag2dc \cite{mag2dc} to investigate the sources of the anomalies observed in the OPM data at 1.8 km along the survey line (Figure~\ref{fig:OPM_Std_plot}) and to interpret the broader magnetic signature of the full profile. Both cross-sections have a bearing of 146° from north, a geomagnetic intensity of 50,363~nT, a magnetic inclination of 70.3$^\circ$ and a declination of -0.8$^\circ$, based on the World Magnetic Model 2025~\cite{wmm2025}. No susceptibility value was assigned to the background rock and we instead assume that the observed magnetic signature is solely due to buried bodies with remanent magnetisation. Note that we do not consider induced magnetisation in the modelling. It is also assumed that the modelled bodies extend a chosen distance along strike and that the cross-section bisects them at 90$^\circ$. \red{It is also assumed that the 3-5 m positional error associated with the GPS-based location has a minimal impact on the forward modelling over the $\sim$200 m span of the modelled anomaly.}\\
\\
In the forward modelling approach, the magnetic anomaly of a given source model is calculated and compared to the observed data. The model parameters (body geometry, depth and magnetic susceptibility) are then iteratively adjusted to improve the fit, which, in this study, was assessed based on visual comparison between the modelled and observed fields. However, due to the inherent non-uniqueness of forward modelling, the resulting model represents only one plausible interpretation of the subsurface from the magnetic data alone, rather than a unique solution.


\bibliography{sample}



\section*{Acknowledgements}
We thank the landowners at Middleton of Glasclune Farm for allowing the magnetic basestation to be situated on their land during the survey. The results presented in this paper rely on data collected at the Eskdalemuir Magnetic Observatory. We thank the British Geological Survey for supporting its operation and INTERMAGNET for promoting high standards of magnetic observatory
practice (www.intermagnet.org). Figures 1 and 2 make use of British Geological Survey survey maps and aeromagnetic survey data respectively, this data is provided under the Open Government License. The PPM instrument used in this survey was provided by the University of Aberdeen (Aberdeen University Geophysical Equipment Repository – AUGER). 

\section*{Funding}
Funding from this survey was provided by the SAGES and SUPA collaboration fund to A.G. and S.I. K.L. acknowledges NERC grant NE/W008289/1. 
A.F. acknowledges the E4 NERC doctoral training partnership grant NE/S007407/1.
S. I. and S. S. acknowledge funding from the UK Quantum Technology Hub in Sensing, Imaging and Timing (QuSIT), EP/Z533166/1. C.D. acknowledges PhD studentship funding from UK MoD.

\section*{Author contributions statement}

All authors contributions (using on CreDIT taxonomy) are outlined below: 
Stirling Scholes: Methodology, Formal analysis, Investigation, Writing - Original Draft, Visualization
Alissa Forsythe: Methodology, Formal analysis, Investigation, Writing - Original Draft, Visualization
Courtney Dyer: Investigation, Writing - Original Draft
Amy Gilligan: Conceptualization, Formal analysis, Investigation, Writing - Original Draft, Funding acquisition
Karen Lythgoe: Conceptualization, Investigation, Writing - Review \& Editing, Supervision, Funding acquisition
Stuart Ingleby: Conceptualization, Methodology, Investigation, Writing - Review \& Editing, Supervision, Funding acquisition
Jenny Jenkins: Investigation, Writing - Original Draft
Marcin Mrozowski: Methodology, Investigation
Jack-Andrew Smith: Investigation

\section*{Additional information}

\textbf{Data availability} The datasets generated during and/or analysed during the current study are available in the University of Strathclyde open access repository, [\url{https://doi.org/10.15129/7aa4f892-e946-4e29-832f-9c8025722adf}]; \textbf{Competing interests} The authors declare no competing interests.

\end{document}